\documentclass{article}
\usepackage{frascatiphys}
\usepackage{graphicx}
\usepackage{amssymb}
\usepackage{graphics}
\usepackage{amsmath}
\usepackage{amsfonts}
\usepackage{bm}
\begin{document}

\title{OPEN PROBLEMS IN GRAVITATIONAL PHYSICS}

\author{S. Capozziello     \\
{\em Dipartimento di Fisica, Universit\'a di Napoli "Federico II", and INFN Sez. di Napoli,}\\
{\em  Via Cinthia, I-80126 - Napoli, Italy,}\\
{\em Gran Sasso Science Institute (INFN), Viale F. Crispi, 7, I-67100, L'  Aquila, Italy.}
 \\
 \\
G. Lambiase       \\
{\em Dipartimento di Fisica "E.R. Caianiello", Universit\'a di Salerno, and INFN Sez. di Napoli} \\
{\em I-84084 Fisciano (Sa), Italy.}
}
\maketitle
\baselineskip=11.6pt
\begin{abstract}
We discuss    some fundamental issues  underlying gravitational physics and point out some of the main shortcomings of Einstein's General Relativity. 
In particular, after taking into account the role of the two main objects of relativistic theories of gravity, i.e. the metric and the connection fields, we consider the possibility that they are not trivially related so that the geodesic structure and the causal structure of the spacetime could be disentangled, as supposed in the Palatini formulation of gravity. 
In this perspective, the  Equivalence Principle,  in its weak and strong  formulations,  can play a fundamental role in discriminating among  competing theories.
The possibility of its violation at quantum level could open new perspectives in gravitational physics and in unification with other interactions. We shortly debate the possibility of equivalence principle measurements by ground-based and space experiments.
\end{abstract}
\baselineskip=14pt

\section{Introduction}

As it is well known, General Relativity (GR) is based on the fundamental assumption that space and
time are entangled into a single spacetime structure assigned on  a Riemannian manifold. 
Being a dynamical  structure, it has to reproduce, in the absence of  gravitational field, the Minkowski spacetime.

Like any  relativistic theory of gravity, GR
has to match some minimal
requirements to be considered  a self-consistent  physical theory.
First of all,  it has to reproduce the Newtonian dynamics in the
weak-energy limit, hence it must be able to explain the astronomical dynamics related to  the orbits of planets and the self-gravitating
structures. Moreover, it  passes some  observational tests in the   Solar
System that constitute its experimental foundation\cite{Will93}. 

However, GR should be able to explain the Galactic dynamics,
taking into account the observed baryonic constituents ({\it e.g.} luminous components as
stars, sub-luminous components as planets, dust and gas),
radiation and Newtonian potential which is, by assumption,
extrapolated to Galactic scales. Besides, it should address the problem of large scale structure
as the  clustering of galaxies.  On cosmological scales, it should address the dynamics of the universe,
which means to reproduce the
cosmological parameters as the expansion rate, the density parameter, and so on, in a sef-consistent way.
Observations and experiments, essentially, probe the standard baryonic matter, the
radiation and an attractive overall interaction, acting at all
scales and depending on distance:  this interaction  is   gravity.

In particular, Einstein's GR is based on four main assumptions. They are
\begin{quote}
The "{\it  Relativity Principle}" -  there is no  preferred
inertial frames, i.e. all frames are good frames for Physics.
\end{quote}
\begin{quote}
The "{\it Equivalence Principle}" -
inertial effects are locally indistinguishable from
gravitational effects (which means the equivalence between the
inertial and the gravitational masses). In other words, any gravitational field can be locally cancelled.
\end{quote}
\begin{quote}
The "{\it  General Covariance Principle}" - field
equations must be " covariant" in form, i.e. 
they must be  invariant in form under the action  of spacetime
diffeomorphisms.
\end{quote}
\begin{quote}
The "{\it Causality Principle}" - each point of space-time has to  admit a universally valid
notion of past, present and future.
\end{quote}

\noindent On these bases, Einstein postulated that, in  a
four-dimensional spacetime manifold, the
gravitational field is described in terms
of the metric tensor field $ds^2 = g_{\mu\nu}dx^{\mu}dx^{\nu}$, with the same signature of
Minkowski metric.  The  metric coefficients have the physical meaning of gravitational potentials. Moreover, he  postulated that
spacetime is curved by the distribution of the energy-matter sources.

The above  principles  require
that the spacetime structure has to be determined by either one or
both of the  two following fields:  a Lorentzian metric $g$ and a linear
connection $\Gamma$, assumed by Einstein  to be torsionless. The metric  $g$  fixes the causal structure of spacetime (the
light cones) as well as its metric relations (clocks and rods);
the connection  $\Gamma$  fixes the free-fall, {\it i.e.} the locally
inertial observers. They have, of course, to satisfy a number of
compatibility relations which amount to require that photons
follow null geodesics of $\Gamma$, so that $\Gamma$ and $g$ can be
independent, {\it a priori}, but constrained, {\it a posteriori},
by some physical restrictions. These, however, do not impose that
$\Gamma$ has necessarily to be the Levi-Civita connection of  $g$ \cite{palatini}.

It should be mentioned, however, that there are many shortcomings of GR, both from
a theoretical point of view (non-rinormalizability, the presence of singularities, and so on),
and from an observational point of view.
The latter indeed clearly shows that  GR is no longer capable of addressing 
 Galactic, extra-galactic and cosmic dynamics, unless  the source side of field equations contains some
 exotic form of matter-energy. These new elusive ingredients are  usually addressed as   {\it dark matter"} and {\it dark
energy}  and constitute up to the $95\%$ of the total cosmological amount of matter-energy\cite{lamb}.

On the other hand, instead of changing the source side of the  Einstein field equations,  one can ask for a "geometrical view" to fit  the missing matter-energy  of the  observed Universe.  In such a case, the dark side could be addressed  by extending GR including more geometric invariants into
 the standard Hilbert - Einstein Action. Such effective Lagrangians can be easily justified at fundamental level by any quantization scheme on curved spacetimes\cite{capozzrev}. 
However, at present stage of the research,  this is
nothing else but a matter of taste,  since no final probe discriminating between dark matter and extended gravity has been found up to now.  
Finally, the bulk of observations that should be considered is so high that an effective Lagrangian or a single particle will be difficult to account for the whole phenomenology at all astrophysical and cosmic scales.

\section{Metric or connections?}

In the GR formulation, Einstein assumed that the metric $g$ of the space-time is the fundamental object to describe gravity.
The connection $\Gamma^\alpha_{\mu\nu}=\Big\{ \begin{array}{c} \alpha \\ \mu\nu \end{array} \Big\}_g$ is constituted by  coefficients
with no dynamics. Only $g$ has
dynamics. This means that  the single object $g$ determines,  at the same time,  the causal structure (light
cones), the measurements (rods and clocks) and the free fall of test particles (geodesics).
Spacetime is therefore a couple $\{{\cal M},g\}$ constituted by a Riemannian manifold and a metric.
Even if it was clear to Einstein that gravity induces  freely falling observers  and that the
Equivalence Principle  selects an object that cannot be a tensor (the connection $\Gamma$) - since it can be  switched off and set to zero at least in a point - he was obliged to
choose it (the Levi - Civita connection) as being determined by the metric structure itself.

In the Palatini formalism a (symmetric) connection $\Gamma$ and a metric $g$ are given and varied independently. Spacetime is a triple
$\{{\cal M}, g, \Gamma\}$ where the metric determines rods and clocks (i.e., it sets the fundamental measurements of spacetime) while $\Gamma$
determines the free fall.
In the Palatini formalism, $\Gamma^\alpha_{\mu\nu}=\Big\{ \begin{array}{c} \alpha \\ \mu\nu \end{array} \Big\}_g$ are differential  equations.
The fact that $\Gamma$ is the Levi-Civita connection of $g$ is no longer an assumption but becomes an outcome of the field equations.

The connection is the gravitational field and, as such, it is the fundamental field in the Lagrangian. The metric $g$ enters the Lagrangian with an "ancillary" role. It reflects the fundamental need  to define lengths and distances, as well as areas and volumes. It defines rods and clocks  that we use to make experiments. It defines also the causal structure of spacetime. However, it has no dynamical role. There is no  whatsoever reason  to assume $g$ to be the potential for $\Gamma$,  nor that it has to be a true field just because it appears in the action.

\section{The role of Equivalence Principle }
The Equivalence Principle (EP) is strictly related to the above considerations and could play a very relevant role in order to discriminate among theories. In particular, it could specify the role of $g$ and $\Gamma$ selecting between  the metric and Palatini formulation of gravity. In particular, precise measurements of EP could say us if $\Gamma$ is only Levi - Civita or a more general connection disentangled, in principle, from $g$.

Let us discuss some topics related to the EP.  Summarizing, the relevance of this principle comes from  the following points:

\begin{itemize}
  \item Competing theories of gravity can be discriminated according to the validity of EP;
  \item EP holds at classical level but  it could be violated at  quantum level;
  \item EP allows to investigate  independently  geodesic and causal structure  of spacetime.
\end{itemize}

\noindent From a theoretical point of view, EP lies at  the physical foundation of metric theories of gravity. The first formulation of EP comes out from the theory of gravitation formulates by Galileo and Newton, i.e. the Weak Equivalence Principle (WEP) which asserts the inertial mass $m_i$ and the gravitational mass  $m_g$ of any physical object are equivalent.
The WEP statement implies that it is impossible to distinguish, locally, between the effects of a gravitational field from those experienced in uniformly accelerated frames using the simple observation of the free falling particles behavior.

A generalization of WEP claims that  Special Relativity is  locally valid. Einstein realized, after the formulation of Special Relativity, that the mass can be reduced to a manifestation of energy and momentum. As a consequence, it is impossible to distinguish between an uniform acceleration and an external gravitational field, not only for free-falling particles,  but whatever is the experiment. According to this observation, 
Einstein EP states:
\begin{itemize}
  \item The WEP is valid.
  \item The outcome of any local non-gravitational test experiment is independent of the velocity of free-falling apparatus.
  \item The outcome of any local non-gravitational test experiment is independent of where and when  it is performed in the universe.
\end{itemize}
One defines as "local non-gravitational experiment" an experiment performed in a small-size of a free-falling laboratory.
Immediately, it is possible to realize  that the gravitational interaction depends on the curvature of spacetime, i.e. the postulates of any metric theory of gravity have to be satisfied.
Hence  the following statements hold:
\begin{itemize}
  \item Spacetime is endowed with a metric $g_{\mu\nu}$.
  \item The world lines of test bodies are geodesics of the metric.
  \item In local freely falling frames, called local Lorentz frames, the non-gravitational laws of physics are those of Special Relativity.
\end{itemize}
One of the predictions of this principle is the gravitational red-shift, experimentally verified by Pound and Rebka in 1960\cite{Will93}.
Notice that gravitational interactions are excluded from WEP and Einstein EP.

In order to classify alternative theories of gravity, the gravitational WEP and the Strong Equivalence Principle (SEP) has to be introduced.
On the other hands, the SEP extends the Einstein EP by including all the laws of physics in its terms. That is:

\begin{itemize}
  \item WEP is valid for self-gravitating bodies as well as for test bodies (gravitational WEP).
  \item The outcome of any local test experiment is independent of the velocity of the free-falling apparatus.
  \item The outcome of any local test experiment is independent of where and when in the universe it is performed.
\end{itemize}
 Alternatively, the Einstein EP is recovered from SEP as soon as the gravitational forces are neglected.
Many authors claim that the only theory coherent with SEP is GR.

A very important issue is the consistency of EP with respect to the Quantum Mechanics.
GR is not the only theory of gravitation and, several alternative theories of gravity have been investigated from
the 60's \cite{capozzrev}.  Considering the spacetime to be special relativistic at a background level, gravitation can be treated  as a Lorentz-invariant field on the background.  Assuming the possibility of GR extensions, two different classes of experiments can be conceived:

\begin{itemize}
  \item Tests for  the foundations of gravitational  theories considering  the various formulations of  EP.
  \item Tests of  metric theories where spacetime is a priori endowed with a metric tensor and where the Einstein EP is assumed always valid.
\end{itemize}
The subtle difference between the two classes of experiments lies on the fact that EP can be postulated a priori or, in a certain sense, "recovered" from the self-consistency of the theory. 
What   is today clear is that, for several fundamental reasons, extra fields are necessary to describe  gravity with respect to the other interactions. Such fields can be   scalar fields or higher-order corrections of curvature invariants\cite{capozzrev}. For these reasons,   two sets of field equations can be considered: The first set couples the gravitational field to the non-gravitational contents of the Universe, i.e. the matter distribution, the electromagnetic fields, etc. The second set of equations gives the evolution of non-gravitational fields. Within the framework of metric theories, these laws depend only on the metric and this is a consequence of the Einstein EP.
In the case where Einstein field equations are modified and matter field are minimally coupled with gravity, we are dealing with the so-called {\it Jordan frame}. In the case where Einstein field equations are preserved and matter field are non-minimally coupled, we are dealing with the so-called {\it Einstein frame}. Both frames are conformally related but the very final issue is to  understand if passing from one frame to the other (and vice versa) is physically significant. 
Clearly, EP plays a fundamental role in this discussion.  In particular, the question is  if it is always valid or  can be violated at quantum level.

\section{Conclusions}

Two of the main challenges of gravitational physics is to test EP in order to discriminate among competing theories of gravity and establish its validity  at quantum level. Different ground-based and space experiments  are devoted to this target that is strictly related to the connection of gravity with other interactions. 
In some sense, considering gravity under the same standard of the other interactions is become a compelling issue after the discovery of Higgs boson at CERN.
For example, very recently a new mission, denoted STE-QUEST (Space-Time Explorer and QUantum Equivalence Principle Space Test) has been proposed. This  mission is designed to answer a range of questions in fundamental physics by performing precision measurements with high accuracy atomic sensors. In particular, STE-QUEST would like to test the Einstein EP, to explore the boundaries with the quantum world, and to search for new fundamental constituents\cite{capozzSTE,STEQUEST}. This will provide a great opportunity for testing fundamental theories of gravity, and, in particular, those theories that extend GR.

\section*{Acknowledgements}

S.C. acknowledges the support of INFN ({\it iniziativa specifica} QGSKY). G. L.  thanks the ASI
(Agenzia Spaziale Italiana) for partial support  (contract ASI  I/034/12/0).

\end{document}